\documentclass[11pt]{article}

\usepackage[iso-8859-7]{inputenc}
\usepackage{epsfig}
\usepackage{graphicx}
\usepackage{epstopdf}
\usepackage{amsfonts}
\usepackage{amssymb}
\usepackage{amsthm}
\usepackage{amsmath}
\usepackage{multirow}
\usepackage{cite}

\usepackage{color}

\newcommand{\nn}{{\nonumber}}

\newcommand{\beq}{\begin{equation}}
\newcommand{\eeq}{\end{equation}}
\newcommand{\bea}{\begin{eqnarray}}
\newcommand{\eea}{\end{eqnarray}}

\newcommand{\gsim}{\lower.7ex\hbox{$\;\stackrel{\textstyle>}{\sim}\;$}}
\newcommand{\lsim}{\lower.7ex\hbox{$\;\stackrel{\textstyle<}{\sim}\;$}}

\newcommand{\be}{\begin{equation}}
\newcommand{\ee}{\end{equation}}
\newcommand{\ba}{\begin{eqnarray}}
\newcommand{\ea}{\end{eqnarray}}

\newcommand{\V}{\mathcal{V}}
\newcommand{\K}{\mathcal{K}}

\usepackage{float}
\usepackage{amsmath} 
\usepackage{amssymb}   
\usepackage{amsthm} 
\usepackage{geometry}
\usepackage{graphicx} 
\usepackage{multirow} 
\numberwithin{equation}{section}

\begin{document} 
	\thispagestyle{empty}
	
	\begin{titlepage}
		
		\vspace*{0.7cm}

		\begin{center}
			{\Large {\bf  Logarithmic loop corrections, 
			moduli stabilisation and de Sitter vacua in string theory
			}}
			\\[12mm]
			Ignatios Antoniadis$^{a,b}$~\footnote{E-mail: \texttt{antoniad@lpthe.jussieu.fr}}, 
			Yifan Chen$^{a,c}$~\footnote{E-mail: \texttt{yifan.chen@lpthe.jussieu.fr}},
			George K. Leontaris$^{d}$~\footnote{E-mail: \texttt{leonta@uoi.gr}}
		\end{center}
		\vspace*{0.50cm}
		\centerline{$^{a}$ \it
			Laboratoire de Physique Th\'eorique et Hautes \'Energies - LPTHE,
		}
		\centerline{\it
			Sorbonne Universit\'e, CNRS, 4 Place Jussieu, 75005 Paris, France}
		\vspace*{0.2cm}
		\centerline{$^{b}$ \it
			Albert Einstein Center, Institute for Theoretical Physics, University of Bern,}
		\centerline{\it
			Sidlerstrasse 5, CH-3012 Bern, Switzerland}
		\vspace*{0.2cm}
		\centerline{$^{c}$ \it
		CAS Key Laboratory of Theoretical Physics, Insitute of Theoretical
Physics,}
\centerline{\it Chinese Academy of Sciences, Beijing 100190, P.R.China}

		\vspace*{0.2cm}
		\centerline{$^{d}$ \it
			Physics Department, University of Ioannina}
		\centerline{\it 45110, Ioannina, 	Greece}
		\vspace*{1.20cm}
		\begin{abstract}
		
We study string loop corrections to the gravity kinetic terms in type IIB compactifications on Calabi-Yau threefolds or their orbifold limits, in the presence of $D7$-branes and orientifold planes. We show that they exhibit in general a logarithmic behaviour in the large volume limit transverse to the $D7$-branes, induced by a localised four-dimensional Einstein-Hilbert action that appears at a lower order in the closed string sector, found in the past. Here, we compute the coefficient of the logarithmic corrections and use them to provide an explicit realisation of a mechanism for K\"ahler moduli stabilisation that we have proposed recently, which does not rely on non-perturbative effects and lead to de Sitter vacua. Our result avoids no-go theorems of perturbative stabilisation due to runaway potentials, in a way similar to the Coleman-Weinberg mechanism, and provides a counter example to one of the swampland conjectures concerning de Sitter vacua in quantum gravity, once string loop effects are taken into account; it thus paves the way for embedding the Standard Model of particle physics and cosmology in string theory.

		\end{abstract}
	\end{titlepage}

 \section{Introduction}

 One of the main challenges in constructing successful particle physics and cosmology models of string theory origin, is the requirement for a stable vacuum with a positive -albeit tiny today- cosmological constant. Despite the theoretical advances in this topic,  there is no conclusive argument that this goal  has been unequivocally achieved. A generic feature of the effective field theories resulting after compactification, is the appearance of a large number of  moduli fields  in the 
 massless spectrum. A primary purpose is thus to obtain a (meta)-stable vacuum where all moduli are fixed and acquire masses in order to avoid long range forces and other undesirable phenomenological features. 
 It has been  realised, however,  that after moduli stabilisation de Sitter (dS) vacua are scarce, if  at all. Focusing on type IIB string in particular, the  scalar potential of the corresponding effective supergravity constructed from the K\"ahler potential and superpotential exhibits a no scale structure  at the  classical level~\cite{Cremmer:1983bf}. Thus, while the complex structure moduli and the dilaton field are fixed in the presence of $3$-form fluxes from supersymmetry conditions imposed via the superpotential~\cite{Frey:2002hf}, K\"ahler moduli fields are not stabilised unless   quantum corrections are taken into account~\cite{Becker:2002nn, Kachru:2003aw, Balasubramanian:2005zx, Conlon:2005ki}.  Moreover, in general the scalar potential  displays often an anti-de Sitter (AdS) minimum,  and therefore, a suitable mechanism is required to provide an appropriate  uplifting  term  which ensures a vacuum with positive energy~\cite{Kachru:2003aw, Balasubramanian:2005zx}.

Motivated by these facts, in a previous work~\cite{Antoniadis:2018hqy}, we studied the quantum corrections arising from 
 a geometric configuration of three intersecting $7$-branes  in the framework of type IIB/F-theory.  The corrections
break the no-scale structure of the K\"ahler potential and generate a non-zero F-term potential for the K\"ahler moduli.
In addition, Fayet-Iliopoulos  D-terms associated with anomalous $U(1)$ symmetries of the intersecting $D7$ branes, are generated.
These  are sufficient to uplift  the  potential,  and generate  a dS minimum with all the  K\"ahler moduli stabilised to their minima. 
  This result is obtained thanks to the fact 
that the aforementioned quantum corrections have a logarithmic dependence on the  moduli associated with 
the co-dimension two volume~\cite{Antoniadis:1998ax}  transverse to the $D7$ brane. 
 Although this type of dependence can be deduced from a simple 
dimensional analysis, yet the  strength and sign of these contributions plays a  decisive role and should be  derived from a direct string calculation.
Therefore, the main objective of the present work is the precise estimate of these corrections leading to the most dominant modifications  of the K\"ahler  potential in the large volume limit. Actually, in type II (and type I) string compactifications, these contributions are induced by the corrections to the Einstein gravity kinetic terms renormalising the effective four-dimensional (4d) Planck mass~\cite{Antoniadis:1996vw, Antoniadis:1997eg}.
 
 In a compactified theory, the large volume limit is expected to give back the higher dimensional theory. In the presence of branes though (or in general localised defects) the limit is more subtle and one has to break up the total volume into pieces along and transverse to the various world-volumes. Still, in the large transverse volume limit, one would expect brane decoupling. However, this is not true in the case of co-dimension two (or one) if the theory has local tadpoles of (effectively) massless bulk states. Indeed, their emission to the bulk from the localised defect leads to infrared divergences due to the effective propagation in two (or one) dimensions. They behave logarithmically with the size of the bulk in the two-dimensional case and linearly in one dimension.  Examples of this property are the threshold corrections to gauge couplings of $D7$ brane gauge kinetic terms~\cite{Antoniadis:1998ax} and the linear dilaton dependence on the size of the eleventh dimension in heterotic M-theory~\cite{Witten:1996mz}. This is not the case for graviton kinetic terms, which are in principle higher dimensional, associated with closed strings that live in ten dimensions. It was found however that type II strings compactified on a non-trivial 6d Calabi-Yau (CY) manifold lead to a 4d Einstein action localised at points where the Euler number is concentrated when taking the large volume limit~\cite{Antoniadis:2002tr}. This correction arises at the string tree-level for smooth manifolds~\cite{Becker:2002nn} (corresponding to a perturbative correction in $\alpha'$~\cite{Grisaru:1986kw}) and at one-loop level in orbifolds~\cite{Antoniadis:1997eg, Antoniadis:2002tr}.

It is now clear that these localised graviton kinetic terms can receive to the next order logarithmic corrections on the size of the volume transverse to $7$-brane (or orientifold) sources localised at distant points from the graviton kinetic terms, due to the emission of closed strings on non-vanishing local tadpoles. Note that consistency of the theory implies only global tadpole cancellation while local tadpoles are generally present except in special configurations of $D7$ branes on top of orientifold planes~\cite{Antoniadis:1998ax}. In this work, we compute these corrections and we show that they are adequate to stabilise the K\"ahler moduli and in particular the total internal volume, obtaining a de Sitter minimum when suitable D-terms from the $D7$-branes are also taken into account.

 The layout of the paper is as follows. In Section 2, we present a  short description of the runaway problem for stabilising the string moduli at weak coupling, related to the generic  behaviour of the scalar potential for the volume modulus (or for the string dilaton) in string theory and discuss possible solutions through quantum corrections. 
 In Section 3, we discuss the appearance of the induced Einstein-Hilbert (EH) term in four dimensions and  the derivation of the one-loop corrections  through graviton scattering. The logarithmic contributions associated with the exchange of Kaluza-Klein (KK) excitations between the  induced graviton vertices and $7$-brane sources are computed in Section 3.2 and are translated in corrections to the K\"ahler potential. In Section 4, we include D-term contributions and describe the minimisation of the scalar potential and the conditions for a dS vacuum. Section 5 contains a summary of our results and some concluding remarks.

\section{Corrections to the K\"ahler potential and runaway moduli behaviour}

The Dine-Seiberg problem \cite{Dine:1985he} is a long-standing question in moduli stabilisation. It concerns the modulus that controls the perturbative expansion, either in $\alpha'$ (the internal volume), or in string loops (the dilaton). In both cases, the assumption is that in the weak coupled limit, the potential at a certain order of the expansion has a monotonic behaviour towards a vanishing value at inifinity corresponding to the free theory. Thus in order to generate a minimum, there should be at least two terms compensating each other, which in general requires the expansion parameter to be of order 1 and the theory is no longer weakly coupled. This argument points out that the vacuum is either strongly coupled or it needs two sectors compensating each other, somehow at weak coupling. For example, in the KKLT model~\cite{Kachru:2003aw}, non-perturbative effects  in the superpotential \cite{Derendinger:1985kk} fix all the  K\"ahler moduli. In the large volume scenario \cite{Balasubramanian:2005zx, Conlon:2005ki} on the other hand, one only uses non-perturbative effects for small volumes which behave like the holes in Swiss cheese. The back-reaction of the small volume together with $\alpha'$-corrections \cite{Kiritsis:1997em, Antoniadis:1997eg, Becker:2002nn} fix the whole volume at exponentially large size.

Here, we want to check this assumption from the bottom-up point of view. One starts with a general form of the K\"ahler potential for a single K\"ahler modulus with perturbative corrections (in supergravity units):
\be
 {\cal K} = -2 \log\, (\tau^{\frac{3}{2}} +  \eta{f[\tau]}),\label{Kaehler}
 \ee
where $\eta$ is the expansion parameter and $f [\tau]$ is a general function of the K\"ahler modulus $\tau$ which breaks the no-scale structure. The term $\eta{f[\tau]}$ is assumed to be much smaller than the volume $\tau^{\frac{3}{2}}$. The corresponding F-term potential for constant superpotential ${\cal W}_0$ is
 \be 
 V_F (\tau) = \frac{\eta \mathcal{W}_0^2}{2\tau^{9/2}} (3 {f[\tau]} - 4 \tau {f'[\tau]} + 4 \tau^2 {f''[\tau]}) + O(\eta^2)\label{Ftermgeneral},
 \ee
 where we omit the higher order terms in the $\eta$-expansion. We see from eq. (\ref{Ftermgeneral}) that the F-term potential naturally splits into three parts. However, if the correction is a power-like function  $f[\tau] = \tau^k$, the three terms above acquire the same form $V_F \propto \eta \tau^{k-9/2} + O(\eta^2)$, thus always being monotonic in the leading order. In the past, all radiative corrections calculated in terms of either the string coupling $g_s$ (at one-loop level) or $\alpha'$ had this form, such as in \cite{Antoniadis:1996vw, Kiritsis:1997em, Antoniadis:1997eg, Becker:2002nn, Antoniadis:2002tr, vonGersdorff:2005bf, Berg:2005ja, Berg:2005yu, Parameswaran:2006jh, Cicoli:2007xp, Berg:2014ama, Haack:2015pbv, Kobayashi:2017zfd, Haack:2018ufg}. 
 
 The observation in \cite{Antoniadis:1998ax} suggests however that in addition to power behaving functions, one could also have corrections with logarithmic  dependence on the moduli
 \be f [\tau] = \log (\tau). \ee
The interesting fact  now is that the F-term potential has two distinct parts, both with the same power dependence but one of the two proportional to $\log (\tau)$~\cite{Antoniadis:2018hqy, Antoniadis:2018ngr}:
  \be 
  V_F = \frac{\eta \mathcal{W}_0^2}{2\tau^{9/2}} (3 \log (\tau) - 8) + O(\eta^2)~.
  \label{VF}
  \ee
 The two terms could compensate each other and lead to a minimum provided that  the coefficient $\eta$ is negative.
 This mechanism is reminiscent of the one with a Coleman-Weinberg potential~\cite{Coleman:1973jx}, offering an alternative solution to the runaway moduli problem, consistent with perturbation theory in the large volume expansion.
 
The next step is to study whether we could get an exponentially large volume. One trivial solution is to insert an extremely small compactification scale parameter $\mu$ inside the correction $f [\tau] = \log (\mu^4 \tau)$.  This is equivalent to adding a constant $\xi$ inside the logarithm of the K\"ahler potential. Indeed, assuming that the volume  upon stabilisation is exponentially large, one can expand the K\"ahler potential in terms of the total volume $\V = \tau^{\frac{3}{2}}$ of the six-dimensional compactification manifold ${{\cal X}_6}$ in the large volume limit:
\be
 {\K} = -2\log\left({\cal V} +\xi +  \eta \log({\cal V}) +  O(\frac{1}{\V})\right) = -2\log\left({\cal V} +  \eta \log({\mu^6 \cal V}) + O(\frac{1}{\V})\right),\label{KahlerpotentialVexpansion}
 \ee	
where $\mu \equiv e^{{\xi}/{6\eta}}$.  In the case of $\eta < 0$, one can show that  there is a minimum of the effective potential (\ref{VF}) in terms of $\V$, which in the large volume limit is:
	\be  
	{\cal V}_{min}=e^{{13}/3}/\mu^6 \quad;\quad V_F^{min}=\frac{\eta \mathcal{W}_0^2}{3{\cal V}_{min}^{3}}~. 
	\label{Vmineta}
	\ee
It follows that in order to make this solution large enough, as required in the large volume expansion, we assumed a priori that $\mu$ has to be exponentially small, which corresponds to the condition
\be 
\xi \gg -\eta > 0\,. 
\label{signsxieta}
\ee
This is again similar to the situation in the Coleman-Weinberg potential, where $\eta$ and $\xi$ correspond to two different parameters/couplings, such as a quartic scalar interaction and a gauge coupling~\cite{Coleman:1973jx}. In the following sections, we shall present an explicit string theory example, realising the above proposal.

 \section{Quantum corrections from graviton scattering}
 
  In the type II superstring action, in addition to the Einstein-Hilbert (EH) term,  there are fourth order corrections in Riemann
 curvature, generated by multi-graviton scattering. Those which are relevant to our discussion, that generate a localised EH action in four dimensions, are of the form~\cite{Green:1997di, Kiritsis:1997em, Antoniadis:1997eg, Russo:1997mk, Becker:2002nn}:
 \be
\int\limits_{M_{10}} \epsilon^{\mu_1\mu_2\dots \mu_8}\epsilon_{\nu_1\nu_2\dots \nu_8}R_{\mu_1\mu_2}^{\nu_1\nu_2} R_{\mu_3\mu_4}^{\nu_3\nu_4} R_{\mu_5\mu_6}^{\nu_5\nu_6} R_{\mu_7\mu_8}^{\nu_7\nu_8}
 \equiv \int\limits_{M_{10}}\epsilon_8\epsilon_8 R^4  = \int\limits_{M_{10}} R\wedge R\wedge R\wedge R\wedge e\wedge e \,,
 \label{t8R4}
 \ee 
where 
the last expression is in differential forms notation, 
$\int_{M_{10}} R^4\wedge e^2$ in short.
The reduction of this term in four dimensions has implications in the effective field theory. Among others, there is an induced localised EH term in four dimensions\cite{Antoniadis:2002tr}   which generates the universal correction to the Planck mass  appearing in eq.~(\ref{KahlerpotentialVexpansion}). In the following subsections, we  first review the result of \cite{Antoniadis:2002tr} which shows that the constant $\xi$ is proportional to the Euler characteristic $\chi$  of the CY manifold ${\cal X}_6$. Then, we  calculate the exchange of closed strings between these localised EH terms and extended object sources ($7$-branes) located at distant points in the bulk. Such local tadpoles lead to logarithmic corrections in the size of the two-dimensional space transverse to the $7$-branes.

 \subsection{Localised Einstein-Hilbert terms}

The implications of $R^4$ terms in the effective theory obtained after
compactifying on a CY manifold, have been studied by several authors
\cite{Antoniadis:1997eg, Kiritsis:1997em, Antoniadis:2002tr, Becker:2002nn}. The term of interest to us
 arising from (\ref{t8R4})  after integration on CY,  is  proportional
to the Euler characteristic of the manifold $\chi$~\cite{Antoniadis:1997eg}. 
Hence, including the tree-level and the one-loop generated Einstein-Hilbert terms, 
after compactification on ${\cal M}_4\times {\cal X}_6$ with ${\cal M}_4$ the 4d Minkowski spacetime, the ten-dimensional action reduces to~\cite{Antoniadis:1997eg, Kiritsis:1997em, Antoniadis:2002tr, Becker:2002nn}
 \ba 
 S &\supset &\frac{1}{(2\pi)^7 \alpha'^4} \int\limits_{M_{10}} e^{-2\phi} {\cal R}_{(10)} - \frac{6}{(2\pi)^7 \alpha'} \int\limits_{M_{10}} \left(-2\zeta(3) e^{-2\phi} \pm 4\zeta(2) \right) R^4 \wedge e^2\label{IIB10DactionA}\\  
 &\equiv&S_{\rm grav}= \frac{1}{(2\pi)^7 \alpha'^4} \int\limits_{M_{4} \times {{\cal X}_6}} e^{-2\phi} {\cal R}_{(10)} - \frac{\chi}{(2\pi)^4 \alpha'} \int\limits_{M_{4}} \left(-2\zeta(3) e^{-2\phi} \pm\frac{2\pi^2}{3}\right) {\cal R}_{(4)}\,,  
 \label{IIB10Daction} 
 \ea
where $\phi$ is the string dilaton whose vacuum expectation value (VEV) defines the string coupling $g_s=\langle e^\phi\rangle$, ${\cal R}_{(d)}$ is the $d$-dimensional Ricci scalar, the $\pm$ sign corresponds to the type IIA/B theory,
and in the second line, $\zeta(2)={\pi^2}/6$ has been  substituted. We also used~\cite{Antoniadis:1997eg}:
\be
\frac{1}{(2\pi)^6}\int\limits_{{\cal X}_6} R\wedge R\wedge R=\frac{\chi}{3!(2\pi)^3}\,.
\ee

Note that the action~(\ref{IIB10Daction}) is universal for smooth manifolds (for orbifolds the tree-level contribution proportional to $\zeta(3)$ vanishes - see below) and the 4d term proportional to $\chi$ corresponds to the large volume limit of the internal compactification space, associated with the localised contributions that remain finite in the non-compact limit. In general, there are extra model dependent terms that vanish at large volume, either perturbatively (power suppressed), or non-perturbatively (exponentially suppressed). The localisation points are arbitrary, corresponding to the singularities supporting the Euler number of the CY manifold in the non-compact limit; thus, in general $\chi=\sum_i\chi_i$, where the summation index $i$ runs over all  different points of localised gravity. In the following, we consider for simplicity that there is just one singularity at the origin supporting the total Euler number. In the orbifold limit, the first term in the bracket involving $\zeta(3)$ vanishes and the localised ${\cal R}_{(4)}$ term arises at one loop, proportional to $\zeta(2)$.
 
From  the above reduction  it is readily inferred that  an induced  term linear in the Ricci scalar is only possible in four 
dimensions~\footnote{Actually, in M-theory gravity localisation arises in five dimensions which is the strongly coupled limit of type IIA.}.
In supersymmetric type II string compactifications, the Euler characteristic  counts the difference between the number of $\mathcal{N} = 2$
 hypermultiplets and vector multiplets,
 \be\chi =\pm 4 (n_V - n_H)\label{chiforII}\,,\ee
 for type IIA/B, respectively.  It is remarkable that
the emergence of a lower dimensional Einstein-Hilbert term ${\cal R}_{(d)}$, with the whole internal space in the non-compact limit, 
is only possible in four dimensions (in the weakly coupled regime) and leads to interesting cosmological consequences that we shall discuss at the end.

\noindent 
To simplify the expression of the action, we choose the mass conventions $2\pi \alpha' = 1$~\cite{Becker:2002nn}, 
leading to:
\ba
S_{\rm grav}&=&\frac{1}{(2\pi)^3} \int\limits_{M_{4} \times {{\cal X}_6}} e^{-2\phi} {\cal R}_{(10)} - \frac{\chi}{(2\pi)^3}  \int\limits_{M_{4}} \left(-2\zeta(3) e^{-2\phi} \pm\frac{2\pi^2}{3}\right) {\cal R}_{(4)}\nonumber\\
&=&\frac{1}{(2\pi)^3} \int\limits_{M_{4}} 
\left[{\cal V} e^{-2\phi} + \chi\left(2\zeta(3) e^{-2\phi} \mp\frac{2\pi^2}{3}\right)\right]  {\cal R}_{(4)}\,.
\label{IIB10DactionF}
\ea
In order to deduce the physical corrections to the moduli metric, one has to take into account also the corrections to the moduli kinetic terms and perform an appropriate Weyl rescaling of the spacetime metric from the string to the Einstein frame where gravity kinetic terms are correctly normalised. It turns out that the corrections to the moduli metric read~\cite{Antoniadis:1997eg}:
\be
{\cal V} e^{-2\phi} \left[(\partial V)^2+(\partial H)^2\right] 
\mp \chi\left(2\zeta(3) e^{-2\phi} + \frac{2\pi^2}{3}\right)\left[(\partial V)^2-(\partial H)^2\right]\,,
\ee
where $V$ and $H$ denote collectively the $\mathcal{N}=2$ vector and hypermultiplet moduli, respectively (orthogonal to the volume and dilaton directions)\footnote{For the volume and universal dilaton hypermultiplet, special care is needed to take care of the mixing~\cite{Antoniadis:2003sw}.}. As a result in type IIA, upon a Weyl rescaling to the Einstein frame, the correction proportional to $\zeta(3)$ renormalises the metric of vector moduli and the dilaton dependence drops, whilst the correction proportional to $\pi^2/3=2\zeta(2)$ renormalises the metric of the hypermultiplet moduli and acquires a dilaton dependence. Thus, the vector and hyper metrics decouple, in agreement with the $\mathcal{N}=2$ supergravity. 

 In type IIB on the other hand, which is our case of interest, it is easy to see 
that in the Einstein frame the vector moduli space is not corrected while the
 hypermultiplet moduli obtain tree-level as well as one-loop corrections. The latter
can be read off from the corrections to ${\cal R}$ up to a factor of $(-2)$ due to the Weyl rescaling. Note that now the internal volume, as well as all K\"ahler class moduli, are in $\mathcal{N}=2$ hypermultiplets, together with the dilaton. After the orientifold projection and turning on $3$-form fluxes, supersymmetry is broken to $\mathcal{N}=1$ keeping only the K\"ahler class and complex structure moduli (and the dilaton) in chiral multiplets. The latter together with the dilaton are stabilised in a supersymmetric way via the $3$-form flux generated superpotential, leaving the K\"ahler moduli unfixed as flat directions when the corrections proportional to $\chi$ are ignored. These corrections modify the tree-level K\"ahler potential by shifting the volume according to eq.~(\ref{KahlerpotentialVexpansion}) by a constant $\xi$ that can be read off from eq.~(\ref{IIB10DactionF}) and taking care of the dilaton fixing following for instance the procedure of~\cite{Becker:2002nn}: $\xi=-\frac{\chi}{4}[\zeta(3)+\pi^3/3]$.\footnote{Note that our $\xi$ differs from the one of~\cite{Becker:2002nn} by a factor of 2.}

Next, we turn to the localisation width of the wavefunction of ${\cal R}_{(4)}$ in the internal manifold ${{\cal X}_6}$ when $\chi \neq  0$. 
In order to render the computations tractable, we work  in the context of type IIB string theory compactified on the $T^6/{Z_N}$ orbifold limit  of CY space ~\cite{Antoniadis:2002tr}. Note that, in the non-compact limit, $N$ can be an arbitrary integer. The non-vanishing contribution to the localised EH action comes  from one-loop, since the tree-level correction (proportional to $\zeta(3)$) vanishes in the orbifold limit~\cite{Antoniadis:1997eg}. Moreover, the one-loop correction receives non-vanishing contributions only from the odd-odd spin structure and from the $\mathcal{N} = 1+1$ supersymmetric sectors with all internal coordinates twisted by the orbifold. In the odd-odd spin structure, one has to take one vertex in the $(-1, -1)$-ghost picture, the two others in $(0, 0)$-ghost picture, and add a world-sheet supercurrent insertion in both left- and right-moving sectors.  

\begin{figure}[h!]
  	\centering
  \includegraphics[width=0.35\columnwidth]{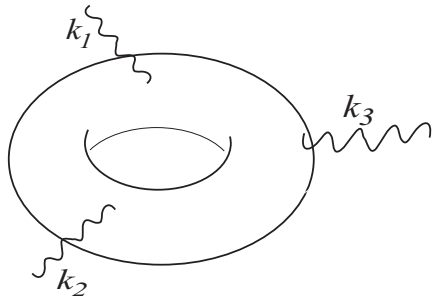}\;\;
  	\caption{
  		\footnotesize
  		{Three graviton scattering with two massless gravitons with momentum $k1, k2$ and a KK state carrying momentum $k_3$.}
  	}
  	\label{xx0}
  \end{figure}

In order to compute the localisation width, the corresponding graviton scattering amplitude involves two massless gravitons and one Kaluza-Klein (KK) excited state with zero winding in the $Z_N$ orbifold background (see Fig.\ref{xx0}). Their 4d momenta satisfy momentum conservation and mass-shell conditions
  $$\sum_i k_i=0,\; k_1^2=k_2^2=0,\;k_3^2=-q^2~,$$
with $q$ being the KK momentum.
  We take the zero modes of the fermions along the non-compact directions from $V_{(0, 0)} (z_1)$ and $V_{(-1, -1)} (z_2)$ and the zero mode parts from the contractions between the supercurrents and the vertex operator $V_{(0, 0)} (z_3)$.  After some manipulations the amplitude takes the form~\cite{Antoniadis:2002tr}
 \ba 
 \langle V^2_{(0,0)}V_{(-1,-1)}\rangle  = {\cal C_R} \frac{1}{N^2} 
 \sum_{\substack{f=0,\dots, n_f\\
       k=0,\dots, N-1}}e^{i \gamma^kq \cdot  {x}_f} \int_{\cal F}\frac{d^2\tau}{\tau_2^2}
   \int \prod_{i=1,2,3}\frac{d^2z_i}{\tau_2} {\sum_{(h,g)}}' e^{\alpha' q^2 F_{(h,g)}(\tau, z_i)}\,,\label{Amp1}
 \ea
where ${\cal C_R}$ contains the linearised tensorial structure of the three gravitons which comes from the expansion of the scalar curvature ${\cal R}_{(4)}$, whilst $x_f$ are the fixed points of the orbifold and $\gamma^k$ is the representation of the action of the orbifold group. The pairs $(h,g)$ label the orbifold sectors corresponding to the boundary conditions $h$ and $g$ around the two cycles of the world-sheet torus, while the prime in the sum excludes the untwisted sector $(0,0)$ which does not contribute to the amplitude because of the fermion zero-modes integration. For simplicity, we consider $N$ prime, so that there are no $\mathcal{N} = 2+2$ supersymmetric sectors that give also vanishing contribution. The factor $1/N^2$ takes into account  the two orbifold projections around the 2 cycles. As usually, the integration over the 2d torus modulus $\tau$ is restricted in the fundamental domain ${\cal F}$ of the $SL(2,\mathbb{Z})$ modular group. The function $F_{(h,g)}(\tau, z_i)$ is computed in~\cite{Antoniadis:2002tr} in terms of the twisted 2-point function and the coupling between two twisted and one untwisted states on the torus.
 
Obviously, the localisation occurs at the orbifold fixed points. Focusing on one of them, say the origin $x_f = 0$, all the others go to infinity in the non-compact limit, while the summation over $k$  gives a factor of $N$.
We now take the Fourier transform with respect to the KK momentum $q$ in all six internal dimensions, in the non-compact limit, using Euclidean signature $q^2<0$. This gives the coefficient of ${\cal R}_{(4)}$ as an integral over the 6d internal position space $y$ of a localisation function $\delta(y)$: 
\ba
 \delta(y) =&& \frac{1}{N} \int_{\cal F}\frac{d^2\tau}{\tau_2^2}
   \int \prod_{i=1,2,3}\frac{d^2z_i}{\tau_2} {\sum_{(h,g)}}' \frac{1}{8\alpha'^3 F_{(h,g)}(\tau, z_i)^3} e^{-\frac{y^2}{4\alpha' F_{(h,g)}(\tau, z_i)}}\nn\\
   \sim&& N \frac{1}{w^6} e^{-\frac{y^2}{2w^2}}\,,
   \label{6Dcoordinatelocal}
   \ea
where we restored $\alpha'\equiv l_s^2$, with $l_s$ the string length.
It is now clear that $\delta(y)$ exhibits a Gaussian profile with respect to the ratio $\frac{y}{w}$ where  the origin of the coordinate  $y$
 is identified with the fixed point and $w$ is an effective width associated with it. The width can be computed in the large $N$ limit by a saddle point analysis, extremising $F_{(h,g)}$. Evaluating the stationary point of $F_{(h,g)}(\tau, z_i)$ the effective width is found to be~\cite{Antoniadis:2002tr}
   \be
    w^2 \simeq \alpha' F_{(h,g)}(\tau, z_i) |_{min} \sim \frac{l_s^2}{N}\,.
    \label{width}
    \ee
The summation over $(h, g)$  
 in eq. (\ref{6Dcoordinatelocal}) leads then to a factor of $N^2$, taken into account in the second line. Notice that the effective width corresponds to the 4d Planck mass in accordance with the field theory arguments of localised gravity~\cite{Dvali:2000hr}.
 
  Looking back at eq. (\ref{IIB10Daction}), we conclude that the one-loop correction can be written as
\be  
 \frac{4\zeta(2)\chi}{(2\pi)^4 \alpha'} \int\limits_{M_{4}\times {{\cal X}_6}}  \frac{1}{(2\pi)^3} \frac{e^{-y^2/(2 w^2)}}{w^6} {\cal R}_{(4)},
   \label{localisedtermorbifold}
   \ee
 where $|\chi| \sim N$ (see eq.~$\ref{chiforII}$) and the effective width $w$ provides the effective ultraviolet cutoff for the graviton KK modes propagating in the bulk.

\subsection{$7$-branes and logarithmic corrections}
We have seen above that one loop corrections in $\mathcal{N} = 1+1$ orbifold compactifications of type II strings generate localised gravity kinetic terms associated with a 4d EH action proportional to the Euler characteristic of the manifold. We are now ready to show that next order corrections  in general display a logarithmic dependence on the size of two dimensional bulk subspaces transverse to distant $7$-brane sources.
Indeed, localised graviton vertices  can emit gravitons and other closed string states in the bulk towards these sources, generating local tadpoles whose existence can be consistent with global tadpole cancellation. 
  \begin{figure}[h!]
	\centering
  	\includegraphics[width=0.35\columnwidth]{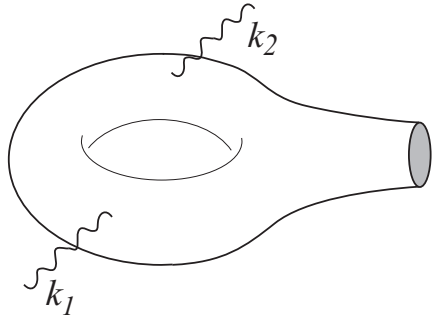}\;\;
	\caption{
		\footnotesize
	{Genus-3/2 amplitude leading to logarithmic correction to the induced 4d Planck mass.}
	}
	\label{xxx}
\end{figure}
The relevant string diagram is genus-3/2, as shown in Fig.~\ref{xxx}, where the disk corresponds to a boundary (brane) or a crosscap (orientifold) and we considered the insertion of two 4d graviton zero-mode vertices.\footnote{As usually, we assume an appropriate off-shell regularisation to prevent on-shell vanishing of the 2-point function. Alternatively, one should consider the $3$-pont function by adding an extra 4d graviton vertex insertion.} The presence of the handle is needed to produce the localised correction in orbifolds, described in the previous subsection, while the presence of the boundary/crosscap is necessary to produce the desired logarithmic correction in the codimension-two case.

The exact computation of this diagram is quite involved. However, the contribution of the local closed string tadpole that gives rise to the logarithmic correction can be done easily in the degeneration limit where the diagram factorises into a $3$-point function on a torus (one loop) of two 4d gravitons and a massless ten-dimensional closed string, and a one-point function on a disk describing the propagation of the closed string in the tube ending on the boundary/crosscap. The massless 10d closed string state can be decomposed in four dimensions into a 4d massless mode and a series of KK excitations for the graviton, dilaton, volume modulus and possibly other (model dependent) untwisted moduli (the 2-index antisymmetric tensor cannot go into the vacuum by Lorentz invariance).

\begin{figure}[h!]
  	\centering
  	   	\includegraphics[width=0.65\columnwidth]{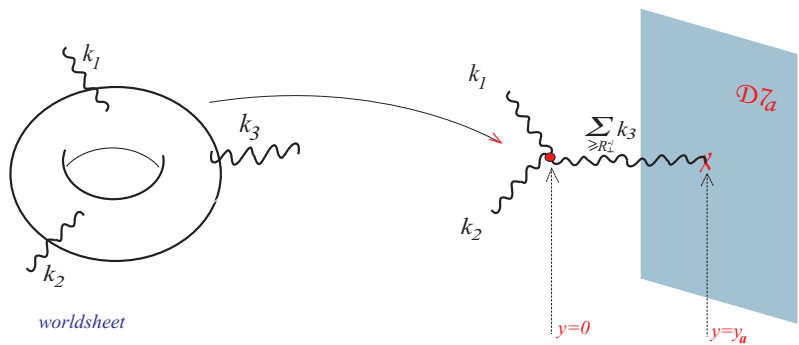}
  	\caption{
  		\footnotesize
{Degeneration limit genus-3/2 amplitude with two massless gravitons and a KK excitation transmitted towards a $D7$ brane.}  	}
  	\label{VinfN=60old}
  \end{figure}

Here, for simplicity, we will use the same one-loop $3$-point function given in eq. (\ref{Amp1}), with a KK mode of the graviton-dilaton (corresponding to the same vertex operator with a different polarisation factor) propagating towards a tadpole ending on a $7$-brane source, see Fig.\ref{VinfN=60old}. 
Obviously, the KK-state must be off-shell with mass $q_\perp$; its 4d momentum is zero by momentum conservation. Similarly, all internal momenta vanish along the $7$-brane world-volume directions. Thus, $q_\perp$ corresponds to the KK-momentum along the two directions transverse to the $7$-brane.
Going back to the momentum space, the $3$-point function~(\ref{6Dcoordinatelocal}) becomes:
\be 
{\tilde\delta}(q_\perp)\sim {\cal C_R} N e^{-w^2 q_\perp^2/2}\,,
 \ee
where ${\cal C_R}$ contains the linearised tensorial structure as in eq. (\ref{Amp1}).

The amplitude  can now be written as the product of the above vertex, the two-dimensional propagator, and the contribution from a $D7$-brane/$O7$-plane. In the string frame, the result is:
\ba
A_S=-{\cal C_R}\sum\limits_{q_\perp\neq 0} g_s^2 T N e^{-w^2 q_\perp^2/2}\frac{1}{q_\perp^2 R_\perp^2 }~,
\label{AS}
\ea 
where $T$ is the brane tension, $R_\perp$ is the size of the two-dimensional  space transverse to the $7$-brane, and the zero mode is omitted from the summation over the KK modes due to the global tadpole cancellation condition.

In the large $R_\perp$ 
limit,  we can  go to the continuum  by replacing the sum with the appropriate integral. 
Thus, including also the Jacobian determinant, we obtain
\ba 
A_S &=&- {\cal C_R} \int_{1/R_\perp}^\infty g_s^2 T N e^{-w^2 p^2/2}\frac{1}{p^2}\frac{2\pi p}{N\sin{\frac{2\pi}{N}}} dp\nn\\
 &=&- {\cal C_R} g_s^2 T \frac{2\pi}{\sin{\frac{2\pi}{N}}}  \frac{1}{2}\Gamma \left(0, \frac{w^2}{2R^2_\perp}\right)\nn\\
 &=&- {\cal C_R} g_s^2 T\frac{2\pi}{\sin{\frac{2\pi}{N}}}  \left\{-\gamma/2 + \log{\left(\frac{R_\perp \sqrt{2}}{w}\right)} + {\cal O}\left(\frac{w^2}{R^2_\perp}\right) \right\}\,,
 \label{ASint}
 \ea
 where $2\pi/(N\sin{\frac{2\pi}{N}})$ is the result of the angular integration corresponding to the volume of $Z_N$ fundamental cell (valid for $N>2$)~\cite{Antoniadis:1993jp}.
Focusing on the $R_\perp$ dependent part, we observe that in the large transverse volume limit, the dominant 
contribution comes from the logarithmic factor, 
\be  
\sim -N g_s^2 T  \log\frac{R_\perp}{w}\,,
 \ee
where we considered also the large $N$ (or $\chi$) limit.
 The above computation can also be done in the position space in a 
 straightforward way, using the localised form factor (\ref{6Dcoordinatelocal}). The result is $\sim N g_s^2 T\log (y_B/w)$, with $y_B$ the distance of the $7$-brane probe from the origin.

Remarkably, the  logarithmic contributions found above, have the opposite sign compared to the one of the constant correction~(\ref{6Dcoordinatelocal}) or (\ref{localisedtermorbifold}). This relative negative sign arises due to the fact that, using for instance Euclidean signature, the propagator ${1}/{q_\perp^2}$ changes sign. Furthermore, the tension $T$ is positive for a $D7$-brane probe and  negative for an $O7$-plane. Thus, the negative value of the parameter $\eta$ (relative to $\xi$) introduced in eq.~(\ref{KahlerpotentialVexpansion}), required to ensure large volume expansion  (see eq. (\ref{Vmineta})), is associated with the existence of local tadpoles through $D7$-branes. This implies that the configuration should involve a surplus of branes relative to orientifold planes. For instance, in the absence of any fluxes, this condition can be satisfied if one places all branes away from the origin (where 4d gravity is localised), at the boundary of the internal space. In the presence of fluxes, this condition can  also be satisfied in several ways.
 Actually, the whole six-dimensional internal space allows at most three different directions of local tadpoles associated with the three possible orthogonal sets of $7$-branes, through the exchange of KK excitations along the corresponding codimension-two transverse dimensions. 

In general, consistent string models should satisfy global tadpole cancellation conditions. Supersymmetric constructions have $3$- and $7$-brane sources (orientifold $O$-planes and $D$-branes) that are subject to global cancellation requirement, where $3$-form fluxes contribute to $3$-form charges. Magnetised $D7$-branes generate also in principle $3$-form charges that should be taken into account~\cite{Antoniadis:2006eu}. Note, however, that $3$-brane sources do not lead to logarithmic corrections, since the corresponding local closed string tadpoles are in a six-dimensional bulk. Thus, only $7$-brane sources give rise to logarithmic corrections with closed strings propagating towards local tadpoles in two dimensions. As an example, their contribution in the $\mathbb{Z}_2$-case reads~\cite{Antoniadis:1998ax}:
 \be
 F(q_\perp)\sim -16\prod_{I=1,2}\frac{1+(-)^{n_I}}{2}+\sum_{a=1}^{16}\cos(q_\perp\cdot y_a)\,, 
 \label{Ftotal}
\ee
where the first term corresponds to the contribution of the $O7$-planes located at the four corners of a square of size $R_\perp$, describing the compactification of the two bulk dimensions, and $y_a$ denote the positions of the $D7$-branes. The function $F(q_\perp)$  normally 
should multiply the terms in the sum of eq.~\eqref{AS} which was done for one $7$-brane source in the bulk at a position of distance $\sim\cal{O}(R_\perp)$ from the origin. The parameter $T$ in \eqref{AS} stands for the total tension of coincident $D7$-branes at this location, as well as of an orientifold if the position is chosen to be at one of the corners of the square away from the origin. The complete computation should include instead the function $F(q_\perp)$. Global tadpole cancellation implies that there are 16 $D7$-branes, cancelling the divergence at $q_\perp=0$ when summing over transverse momenta $q_\perp$. It is then easy to see that for generic positions of the branes away from the origin $y_a= c_a \pi R_\perp$ with $c_a<1$ but fixed as $R_\perp\to\infty$, the sum in eq.~\eqref{AS} approximated by the integral in eq.~\eqref{ASint} behaves logarithmically in $R_\perp$ with a coefficient $T$ given by the tension of a single $D7$-brane corresponding to the sum of the positive tensions of 16 $D7$-banes and the negative tensions of 15 $O7$-planes (since one of them is located at the origin). The only possibility to cancel the coefficient of the logarithm is to place one $D7$-brane at the origin, where 4d gravity is localised. When more $7$-branes are at the origin the coefficient of the logarithm changes sign. It is then clear that the logarithmic corrections we consider are generic to any globally consistent type IIB compactification with $D7$-branes.

We can now substitute both the one-loop correction $\xi$ and the three genus-3/2 corrections proportional to $\eta \log{R_\perp}$ in the induced action to  obtain
\be
 \frac{1}{(2\pi)^3} \int\limits_{M_{4} \times {{\cal X}_6}} e^{-2\phi} {\cal R}_{(10)} + \frac{4\zeta(2) \chi }{(2\pi)^3} \int\limits_{M_{4}} \left( 1 - \sum_{i = 1, 2, 3} e^{2\phi} T_i  \log\frac{R_\perp^i}{w} \right) {\cal R}_{(4)}~,
 \label{kinterms}
 \ee
where $i$ labels the three transverse directions for each one of the three $D7$-branes, and we restored the Euler number $\chi$ from $N$. We also fixed the normalisation factor ${\cal C_R}$ in the previous equations by the correct coefficient of the one-loop correction to the localised gravity kinetic term ${\cal R}_{(4)}$. From eqs.~(\ref{kinterms}) and~(\ref{IIB10Daction}) (for the smooth CY case), by comparing the 10d and 4d gravity kinetic terms, one obtains in the weak coupling limit:
\be
\xi=-\frac{1}{4}\chi f(g_s)\quad;\quad f(g_s)=
\begin{cases}
\zeta(3)\simeq 1.2\quad {\rm for\ smooth\ CY}\\[3pt]
\frac{\pi^2}{3}g_s^2\quad\hskip 0.9cm {\rm for\ orbifolds}
\end{cases}
\label{cases}
\ee
implying a negative Euler number $\chi<0$, in order to satisfy the condition~(\ref{signsxieta}).

We can now estimate the compactification scale $1/\mu$ introduced in eq.~(\ref{KahlerpotentialVexpansion}). Assuming for simplicity a universal $D7$ brane tension $T=e^{-\phi}T_0$, we get
\be 
\eta=-\frac{1}{2} g_sT_0\xi \quad;\quad  \mu = \frac{1}{w} e^{\frac{\xi}{6 \eta}}  = \sqrt{|\chi|} e^{-\frac{1}{3 g_sT_0}}\,,
\label{xiovereta}
\ee
where the factor $|\chi| \sim N$ comes from the width $w$ in eq.~\eqref{width} entering as an effective ultraviolet cutoff in the argument of the logarithm (see eq.~\eqref{kinterms}).
Thus, by lowering the string coupling (or the brane tension $T_0$), the volume would go exponentially large as desired. As seen above, we  also need $\xi>0$ which implies positive induced Planck mass, requiring a surplus of vectors from the twisted orbifold sectors~\cite{Antoniadis:2002tr}.
Note that $T_0$ is an effective tension depending on the complex structure moduli (in principle fixed by $3$-form fluxes) and internal magnetic fields along the four compactified dimensions of the $D7$-brane world-volume.

We also notice that our result of the logarithmic correction is in principle valid for the case of a general CY compactification, where localised gravity kinetic terms arise at string tree-level due to $\alpha'$-corrections. Indeed, the computation in the degeneration limit we described above is expected to go through. The logarithmic correction has an extra factor of the string coupling $g_s$ relative to $\xi$, since it arises at the next order in the presence of a boundary/crosscap, and has the opposite sign for the same reason explained above. Thus, Equation~(\ref{xiovereta}) should hold in the general case, in the limit of large size of the volume transverse to the $7$-brane source.

 One might  also worry whether  large  back-reaction effects are induced from $7$-branes that are in principle taken into account within F-theory. However such effects are expected to be important at strong coupling or in the presence of scalar VEVs that take the theory away from the orientifold description. Since we do not consider such VEVs and our stabilisation mechanism works at weak coupling, we do not expect that back-reaction effects would be important. Moreover, the emergence of logarithmic corrections is based on infrared effects due to local tadpoles of (effectively) massless states that propagate in two dimensions emitted from localised vertices towards $7$-brane sources. The  latter  exist also in F-theory. One may wonder  whether 4d localised gravity kinetic terms arise in F-theory, as well. Actually, these corrections are present in $N=2$ type IIB compactifications already in the absence of fluxes, orientifolds and $7$-branes, where arguments based on S-duality can be used to establish their presence also in strong coupling~\cite{Antoniadis:2002tr}. It is therefore plausible that they are also present in F-theory, although  we don't discuss them here since this work focuses on perturbative corrections.

\section{D-terms and de Sitter vacua}

In the previous section we established that the F-term potential~(\ref{VF}) has an AdS minimum~(\ref{Vmineta}) with respect to the total volume at a value which becomes exponentially large in the weak coupling limit. This vacuum breaks supersymmetry since for constant superpotential 
the F-auxiliary component of the volume modulus superfield  does not vanish, unlike KKLT~\cite{Kachru:2003aw} but similar to the large volume scenario (although for a different reason)~\cite{Balasubramanian:2005zx, Conlon:2005ki}.
In~\cite{Antoniadis:2018hqy}, we have shown that this minimum can be uplifted to positive energy when appropriate D-term contributions from the $D7$-branes are taken into account. Here, we will analyse again these contributions and make a quantitative argument for the existence of perturbative dS vacua at large volume and weak string coupling.

One way to obtain D-term contributions is by introducing magnetic fluxes of $U(1)$ gauge group factors along the world-volume directions of the $D7$-branes. The corresponding Fayet-Iliopoulos (FI) contribution is:
\bea 
V_{D_i} = \frac{d_i}{\tau_i} \left(\frac{\partial K}{\partial_{\tau_i}}\right)^2 \,=\, 
\frac{d_i}{\tau_i^3} + {\cal O}(\eta_j)\,,
\label{VD-termlimit}
\eea
where $i$ denotes a $D7$-brane stack, $\tau_i$ is the corresponding world-volume modulus and  
the constant $d_i$ is positive and proportional to the magnetic flux.
The lowest order approximation on the right hand side of the equation above is valid in the large volume expansion of the K\"ahler potential (\ref{KahlerpotentialVexpansion}), i.e., when ${\cal V} \gg \sum_j|\eta_j|\ln ({\tau_j^{3/2}}\mu^6)$. 

Considering now three such orthogonal sets of magnetised $D7$-branes, one obtains the total scalar potential as a sum of the F-term~(\ref{VF}) 
and all the D-term contributions~(\ref{VD-termlimit}):
\beq 
\label{Vtotal}
V_{tot} = \frac{3\eta \mathcal{W}_0^2}{\mathcal{V}^3} \left(\textrm{ln} (\mathcal{V}\mu^6) - 4\right) + \frac{d_1}{\tau_1^3} + \frac{d_2}{\tau_2^3} + \frac{d_3 \tau_1^3 \tau_2^3}{\mathcal{V}^6}\,,
\eeq
where we used ${\cal V}=(\tau_1\tau_2\tau_3)^{1/2}$ and we considered for simplicity the case of equal $\eta_i\equiv\eta$, which is not a necessary condition for a global minimum once D-terms are included for all three mutually orthogonal $D7$-brane stacks~\cite{Antoniadis:2018hqy}. 
In deriving the F-term contribution to the scalar potential, special care is needed for the dilaton dependence of the K\"ahler potential that enters non-trivially together with the volume and the other K\"ahler moduli which, in type IIB, all descend  from ${\cal N}=2$ hypermultiplets. This leads to a mixing in the K\"ahler metric between K\"ahler moduli and the dilaton, bringing  an ambiguity in the expression of the scalar potential depending on whether one treats the dilaton as constant before or after applying the supergravity formula~\cite{Becker:2002nn}. However, since the mixing term is either proportional to $\xi$ or $\eta$, the leading order in the large volume expansion used in eq.~\eqref{Vtotal} is not affected from this ambiguity.

\begin{figure}[h!]
 		\centering
 		\includegraphics[width=0.8\columnwidth]{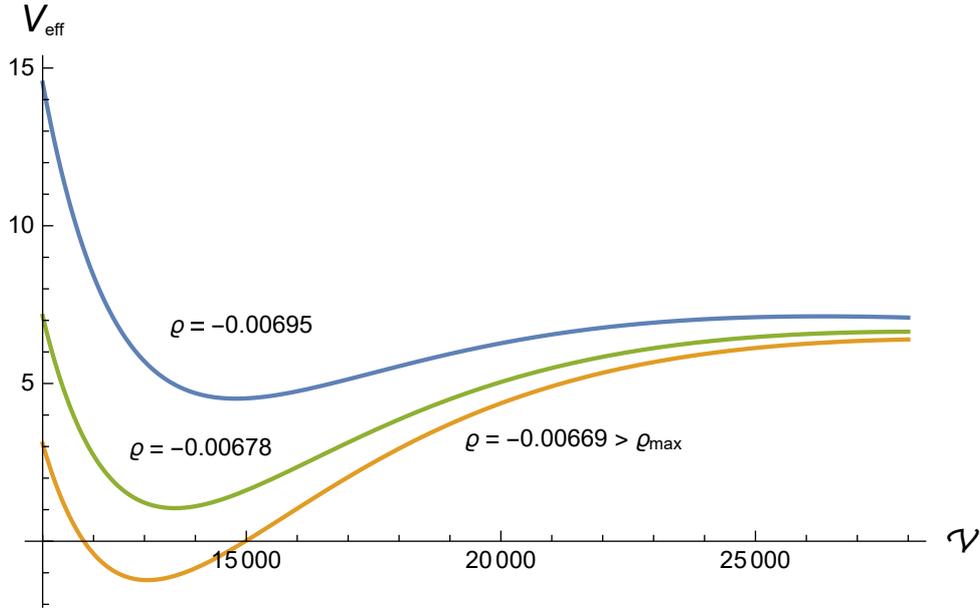}
 		\caption{
 			\footnotesize
 			{Plot of $V_{\rm eff}$  vs ${\cal V}$ (in arbitrary units) for three values of the parameter $\varrho=\frac{d}{\eta \mu^6 {\cal W}_0^2}$.
 				The lower curve corresponds to  AdS  vacuum.
 				 At large ${\cal V}$, $V_{\rm eff}$  vanishes asymptotically after passing from a maximum.}  	}
 		\label{VeffdS}
 	\end{figure}

Minimising the scalar potential \eqref{Vtotal} with respect to $\tau_1$ and $\tau_2$, one fixes the ratios:
\beq
\frac{\tau_i}{\tau_j}=\left(\frac{d_i}{d_j}\right)^{1/3}~,
\eeq
and the scalar potential becomes:
\beq 
V_{tot} = \frac{3\eta \mathcal{W}_0^2}{\mathcal{V}^3} \left(\textrm{ln} (\mathcal{V}\mu^6) - 4\right) + 3\frac{d}{{\cal V}^2}\quad;\quad 
d=(d_1d_2d_3)^{1/3}\,.
\eeq
Minimising now with respect to the volume, one gets:
\beq 
\eta \mathcal{W}_0^2 (13 - 3 \textrm{ln} (\mathcal{V}\mu^6)) = 2d\, \mathcal{V}\,. 
\label{minimizationofV}
\eeq
Following the steps of the analysis of Ref.~\cite{Antoniadis:2018hqy}, one finds that the potential $V_{tot}$ has a dS minimum provided the following inequality is satisfied:
\beq 
-0.007242 < \frac{d}{\eta \mathcal{W}_0^2\mu^6}\equiv\varrho  < -0.006738\,.
\label{inequalitiesDS}
\eeq
The inequality on the left  implies that eq.~(\ref{minimizationofV}) has two solutions with the smaller one being a minimum and the larger one a maximum, so that the potential vanishes asymptotically from positive values. The  inequality on  the right implies that the minimum has positive energy, see Fig.\ref{VeffdS}.
Within this range, the value of the volume at the minimum 
is approximately given by:
\beq 
\textrm{ln} (\mu^6\mathcal{V}_{min}) \simeq 5~,
\label{V0min}
\eeq
and the potential at the minimum reads
\beq
V_{tot}^{min} = \frac{3 \eta \mathcal{W}_0^2}{\mathcal{V}_{min}^3} + \frac{3 d}{\mathcal{V}_{min}^2} > 0\,.
\eeq

Let us now examine what conditions must be imposed  on the parameter space in order to satisfy the inequality (\ref{inequalitiesDS}) at weak string coupling and large volume. We have seen already from eq.~(\ref{xiovereta}) that in the limit $g_s\to 0$, $\mu$ becomes exponentially suppressed and $\mathcal{V}_{min}$ becomes exponentially large. This makes the condition (\ref{inequalitiesDS}) difficult to satisfy, unless $d/(\eta\mathcal{W}_0^2\mu^6)$ also vanishes, compensating the vanishing of $\mu$.
Note that $d$ is proportional to the square of the $U(1)$ gauge coupling, and thus to $g_s$, as well as to the magnetic flux. On the other hand, the brane tension $T_0$ is also proportional to the magnetic flux. It follows from the expression~(\ref{xiovereta}) of $\eta$, that the ratio $\varrho$ defined in
(\ref{inequalitiesDS}) becomes 
\beq
\varrho=\frac{d}{\eta \mathcal{W}_0^2\mu^6}\propto -\frac{1}{\xi\mathcal{W}_0^2\mu^6}\, .
\eeq
Note that ${\cal W}_0$ has been redefined appropriately to be invariant under K\"ahler transformations, taking into account various dilaton dependent and constant factors. In particular, ${\cal W}_0^2$ contains an implicit proportionality factor of $g_s^{-1/2}$~\cite{Becker:2002nn}.

The parameter $\xi$ is given in eq.~(\ref{cases}) and is of order $g_s^2$ for orbifolds and $g_s$-independent for smooth CY manifolds, but in both cases is proportional to the Euler number $\chi$. Thus, in order to satisfy (\ref{inequalitiesDS}) at weak coupling, one has to consider either large $\chi$ or large $\mathcal{W}_0$:
\beq
|\chi| \mathcal{W}_0^2\mu^6 \simeq 100\, .
\label{cond}
\eeq
Large $\chi$ enhances the strength of the localised 4d gravity kinetic term, while large $\mathcal{W}_0$ implies in general large quantised flux.
Both are in principle easy to satisfy. 

As an illustration, in Fig.\ref{VeffdS}, the potential is plotted for $\eta \approx -0.4$, $\varrho\mu^6 \approx 7.5\times 10^{-5}$ and three different values of $\varrho$ larger than the lower bound of eq.~(\ref{inequalitiesDS}), showing how the minimum passes from negative to positive energy values.
One can also check the validity of the approximation (\ref{VD-termlimit}), using (\ref{V0min}) and (\ref{cond}):
\beq
{\cal V} \gg |\eta|\ln ({\cal V}\mu^6)\,\, \Leftrightarrow\,\, \frac{e^5}{\mu^6} \gg 5|\eta| \, ,
\eeq
implying $e^5 \gg 50 g_sT_0/\mathcal{W}_0^2$ that can be easily satisfied for weak coupling and  large $\mathcal{W}_0$.

\section{Conclusions}

Constructing de Sitter vacua within the framework of the effective supergravity of string compactifications is one of the most challenging tasks. 
As it is well known since long time ago, quantum corrections play a pivotal role in accomplishing this goal. In the 
present work we have taken a step forward, by computing higher loop corrections to the Planck mass (and thus to the K\"ahler potential)  in a type IIB
background,  assuming a geometric configuration consisting of three intersecting $D7$-branes.  Our computations rely 
on previous studies where the implications of $R^4$ terms of the ten-dimensional action play an important role.
As shown in ~\cite{Antoniadis:2002tr}, in particular, compactifying on a CY manifold, a localised Einstein-Hilbert  term 
is generated in four dimensions (in the non-compact limit)  which induces a universal correction to the Planck mass multiplied by 
a constant factor proportional to   the Euler characteristic $\chi$.
Furthermore, by studying loop corrections, we find that new non-vanishing contributions are induced by the emission of local tadpoles of closed strings from
the localised gravity  vertices, which exhibit a 
logarithmic dependence on the large co-dimension two volume transverse to each distant $7$-brane probe.
Our computation shows that these corrections, together with the usual D-term contributions from the $D7$-branes world-volume,  suffice to consolidate  a de Sitter vacuum in  type IIB constructions 
based on geometric configurations of the aforementioned type. 

Recently, there have been several arguments casting doubts on the validity of constructing de Sitter vacua using 4d effective supergravity description starting from the no-scale structure, for example in \cite{Sethi:2017phn, Danielsson:2018ztv}, which led to the swampland dS conjecture~\cite{Obied:2018sgi}. Most arguments are related to the validity of the non-perturbative effects, the addition of the anti-$D3$ branes and the 
accuracy of quantum corrections. In this work, only perturbative corrections are invoked based on the structure of the localised 4d Einstein Hilbert term, which is universal and  dependent only on the internal topology. The logarithmic correction is also 
a generic consequence of  
infrared divergences due to local tadpoles of effectively massless closed strings emitted by the localised vertices and  propagating in a two-dimensional bulk.  Its effect in the minimisation of the scalar potential is very important for invalidating the assumptions of the previous arguments, allowing for the presence of locally stable dS vacua in accordance with large volume and weak string coupling, thus providing an explicit counter example to the dS swampland conjecture.

It is important to emphasise that  these corrections exist only in four dimensions thanks to the presence of the induced localised Einstein-Hilbert term. Thus, based on this observation, one can also argue that this mechanism could be used to explain why the Universe is four-dimensional from a new emergent point of view which differs from previous arguments, see for instance~\cite{Brandenberger:1988aj}. More precisely,  one  starts from the ten-dimensional non-compact limit while each   $D7$ brane can compactify the  two dimensions transverse to it. However, the stabilisation mechanism requires a localised source for the graviton kinetic terms which only exists in four dimensions.
This generates local tadpoles of massless closed strings emitted to at most three mutually orthogonal sets of $D7$ branes, which then compactify all six-dimensions. The three intersecting $D7$ branes configuration is also the basic ingredient of type IIB and F-theory models providing an interesting framework for realising the Standard Model of particle physics, see for instance~\cite{Vafa:1996xn, Beasley:2008dc}.

 \section*{Acknowledgements} Y.C. would like to thank Karim Benakli, Peng Cheng, Michele Cicoli, Joseph Conlon, Laura Covi, Jean-Pierre Derendinger, Xin Gao, Mark Goodsell, 	
Mariana Gra\~na, Michael Haack, Tailin Li, Ruben Minasian, Fernando Quevedo, Pramod Shukla and Alexander Westphal for useful discussions. G.K.L. would like to thank the Albert Einstein Center for Fundamental Physics of the University of Bern, and the LPTHE of Sorbonne University for hospitality during various stages of this work. This work was supported in part by the Labex ``Institut Lagrange de Paris'', in part by the Swiss National Science Foundation, in part by a CNRS PICS grant, and in part by the Erasmus exchange program.



  \newpage


\begin{thebibliography}{99}	

\bibitem{Cremmer:1983bf}
  E.~Cremmer, S.~Ferrara, C.~Kounnas and D.~V.~Nanopoulos,
  ``Naturally Vanishing Cosmological Constant in N=1 Supergravity,''
  Phys.\ Lett.\  {\bf 133B} (1983) 61;
  J.~R.~Ellis, C.~Kounnas and D.~V.~Nanopoulos,
  ``Phenomenological SU(1,1) Supergravity,''
  Nucl.\ Phys.\ B {\bf 241} (1984) 406.



\bibitem{Frey:2002hf}
A.~R.~Frey and J.~Polchinski,
``N = 3 warped compactifications,''
Phys.\ Rev.\ D {\bf 65} (2002) 126009
[arXiv:hep-th/0201029];
S.~Kachru, M.~B.~Schulz and S.~Trivedi,
``Moduli stabilization from fluxes in a simple IIB orientifold,''
JHEP {\bf 0310} (2003) 007
[arXiv:hep-th/0201028].
 
	\bibitem{Becker:2002nn}
	K.~Becker, M.~Becker, M.~Haack and J.~Louis,
	``Supersymmetry breaking and alpha-prime corrections to flux induced potentials,''
	JHEP {\bf 0206} (2002) 060
	[hep-th/0204254].
		
	\bibitem{Kachru:2003aw}
	S.~Kachru, R.~Kallosh, A.~D.~Linde and S.~P.~Trivedi,
	``De Sitter vacua in string theory,''
	Phys.\ Rev.\ D {\bf 68} (2003) 046005
	[hep-th/0301240].

\bibitem{Balasubramanian:2005zx}
V.~Balasubramanian, P.~Berglund, J.~P.~Conlon and F.~Quevedo,
``Systematics of moduli stabilisation in Calabi-Yau flux compactifications,''
JHEP {\bf 0503} (2005) 007
[hep-th/0502058].

	\bibitem{Conlon:2005ki}
	J.~P.~Conlon, F.~Quevedo and K.~Suruliz,
	``Large-volume flux compactifications: Moduli spectrum and D3/D7 soft supersymmetry breaking,''
	JHEP {\bf 0508} (2005) 007
	[hep-th/0505076].
	
\bibitem{Antoniadis:2018hqy} 
  I.~Antoniadis, Y.~Chen and G.~K.~Leontaris,
  ``Perturbative moduli stabilisation in type IIB/F-theory framework,''
  Eur.\ Phys.\ J.\ C {\bf 78}, no. 9, 766 (2018)
  [arXiv:1803.08941 [hep-th]].
	
\bibitem{Antoniadis:1998ax}
  I.~Antoniadis and C.~Bachas,
  ``Branes and the gauge hierarchy,''
  Phys.\ Lett.\ B {\bf 450} (1999) 83
  [hep-th/9812093].
	
\bibitem{Antoniadis:1996vw}
  I.~Antoniadis, C.~Bachas, C.~Fabre, H.~Partouche and T.~R.~Taylor,
  ``Aspects of type I - type II - heterotic triality in four-dimensions,''
  Nucl.\ Phys.\ B {\bf 489} (1997) 160
  [hep-th/9608012]. 

	\bibitem{Antoniadis:1997eg}
	I.~Antoniadis, S.~Ferrara, R.~Minasian and K.~S.~Narain,
	``$R^4$ couplings in M and type II theories on Calabi-Yau spaces,''
	Nucl.\ Phys.\ B {\bf 507} (1997) 571
	[hep-th/9707013].

\bibitem{Witten:1996mz}
  E.~Witten,
  ``Strong coupling expansion of Calabi-Yau compactification,''
  Nucl.\ Phys.\ B {\bf 471} (1996) 135
  [hep-th/9602070].
 	
 	\bibitem{Antoniadis:2002tr}
 	I.~Antoniadis, R.~Minasian and P.~Vanhove,
 	``Noncompact Calabi-Yau manifolds and localized gravity,''
 	Nucl.\ Phys.\ B {\bf 648} (2003) 69
 	[hep-th/0209030].
 
\bibitem{Grisaru:1986kw}
  M.~T.~Grisaru, A.~E.~M.~van de Ven and D.~Zanon,
  ``Four Loop Divergences for the N=1 Supersymmetric Nonlinear Sigma Model in Two-Dimensions,''
  Nucl.\ Phys.\ B {\bf 277} (1986) 409.

 
 
 
\bibitem{Dine:1985he} 
  M.~Dine and N.~Seiberg,
  ``Is the Superstring Weakly Coupled?,''
  Phys.\ Lett.\  {\bf 162B}, 299 (1985).
	
\bibitem{Derendinger:1985kk}
  J.~P.~Derendinger, L.~E.~Ibanez and H.~P.~Nilles,
  ``On the Low-Energy d = 4, N=1 Supergravity Theory Extracted from the d = 10, N=1 Superstring,''
  Phys.\ Lett.\ B {\bf 155} (1985) 65;
  ``On the Low-Energy Limit of Superstring Theories,''
  Nucl.\ Phys.\ B {\bf 267} (1986) 365.

	
	\bibitem{Kiritsis:1997em} 
	E.~Kiritsis and B.~Pioline,
	``On $R^4$ threshold corrections in IIb string theory and (p, q) string instantons,''
	Nucl.\ Phys.\ B {\bf 508}, 509 (1997)
	[hep-th/9707018].

 	
	\bibitem{vonGersdorff:2005bf} 
  G.~von Gersdorff and A.~Hebecker,
  ``K\"ahler corrections for the volume modulus of flux compactifications,''
  Phys.\ Lett.\ B {\bf 624}, 270 (2005)
  [hep-th/0507131].


	\bibitem{Berg:2005ja} 
	M.~Berg, M.~Haack and B.~K\"ors,
	``String loop corrections to K\"ahler potentials in orientifolds,''
	JHEP {\bf 0511}, 030 (2005)
	[hep-th/0508043].

\bibitem{Berg:2005yu} 
  M.~Berg, M.~Haack and B.~K\"ors,
  ``On volume stabilization by quantum corrections,''
  Phys.\ Rev.\ Lett.\  {\bf 96}, 021601 (2006)
  [hep-th/0508171].

\bibitem{Parameswaran:2006jh} 
  S.~L.~Parameswaran and A.~Westphal,
  ``de Sitter string vacua from perturbative K\"ahler corrections and consistent D-terms,''
  JHEP {\bf 0610}, 079 (2006)
  [hep-th/0602253].

\bibitem{Cicoli:2007xp} 
  M.~Cicoli, J.~P.~Conlon and F.~Quevedo,
  ``Systematics of String Loop Corrections in Type IIB Calabi-Yau Flux Compactifications,''
  JHEP {\bf 0801}, 052 (2008)
  [arXiv:0708.1873 [hep-th]].
\bibitem{Berg:2014ama} 
  M.~Berg, M.~Haack, J.~U.~Kang and S.~Sj\"ors,
  ``Towards the one-loop K\"ahler metric of Calabi-Yau orientifolds,''
  JHEP {\bf 1412}, 077 (2014)
  [arXiv:1407.0027 [hep-th]].
	
\bibitem{Haack:2015pbv} 
  M.~Haack and J.~U.~Kang,
  ``One-loop Einstein-Hilbert term in minimally supersymmetric type IIB orientifolds,''
  JHEP {\bf 1602}, 160 (2016)
  [arXiv:1511.03957 [hep-th]].
  
\bibitem{Kobayashi:2017zfd} 
  T.~Kobayashi, N.~Omoto, H.~Otsuka and T.~H.~Tatsuishi,
  ``Radiative K\"ahler moduli stabilization,''
  Phys.\ Rev.\ D {\bf 97}, no. 10, 106006 (2018)
  [arXiv:1711.10274 [hep-th]].
  

  \bibitem{Haack:2018ufg}
  M.~Haack and J.~U.~Kang,
  ``Field redefinitions and K\"ahler potential in string theory at 1-loop,''
  JHEP {\bf 1808} (2018) 019
  [arXiv:1805.00817 [hep-th]].



  \bibitem{Antoniadis:2018ngr}
  I.~Antoniadis, Y.~Chen and G.~K.~Leontaris,
  ``Inflation from the internal volume in type IIB/F-theory compactification,''
  Int.\ J.\ Mod.\ Phys.\ A {\bf 34} (2019) no.08,  1950042
  [arXiv:1810.05060 [hep-th]].


\bibitem{Coleman:1973jx}
  S.~R.~Coleman and E.~J.~Weinberg,
  ``Radiative Corrections as the Origin of Spontaneous Symmetry Breaking,''
  Phys.\ Rev.\ D {\bf 7} (1973) 1888.



\bibitem{Green:1997di}
M.~B.~Green and P.~Vanhove,
``D instantons, strings and M theory,''
Phys.\ Lett.\ B {\bf 408} (1997) 122
[hep-th/9704145].

\bibitem{Russo:1997mk}
  J.~G.~Russo and A.~A.~Tseytlin,
  ``One loop four graviton amplitude in eleven-dimensional supergravity,''
  Nucl.\ Phys.\ B {\bf 508} (1997) 245
  [hep-th/9707134].

\bibitem{Antoniadis:2003sw}
  I.~Antoniadis, R.~Minasian, S.~Theisen and P.~Vanhove,
  ``String loop corrections to the universal hypermultiplet,''
  Class.\ Quant.\ Grav.\  {\bf 20} (2003) 5079
  [hep-th/0307268].  
  
\bibitem{Dvali:2000hr}
G.~R.~Dvali, G.~Gabadadze and M.~Porrati,
``4-D gravity on a brane in 5-D Minkowski space,''
Phys.\ Lett.\ B {\bf 485} (2000) 208
[hep-th/0005016];
  G.~R.~Dvali and G.~Gabadadze,
  ``Gravity on a brane in infinite volume extra space,''
  Phys.\ Rev.\ D {\bf 63} (2001) 065007
  [hep-th/0008054].

\bibitem{Antoniadis:1993jp}
See e.g. I.~Antoniadis and K.~Benakli,
  ``Limits on extra dimensions in orbifold compactifications of superstrings,''
  Phys.\ Lett.\ B {\bf 326} (1994) 69
  [hep-th/9310151].

\bibitem{Antoniadis:2006eu}
  I.~Antoniadis, A.~Kumar and T.~Maillard,
  ``Magnetic fluxes and moduli stabilization,''
  Nucl.\ Phys.\ B {\bf 767} (2007) 139
  [hep-th/0610246].

\bibitem{Sethi:2017phn} 
  S.~Sethi,
  ``Supersymmetry Breaking by Fluxes,''
  JHEP {\bf 1810}, 022 (2018)
  [arXiv:1709.03554 [hep-th]].
  
\bibitem{Danielsson:2018ztv} 
  U.~H.~Danielsson and T.~Van Riet,
  ``What if string theory has no de Sitter vacua?,''
  Int.\ J.\ Mod.\ Phys.\ D {\bf 27}, no. 12, 1830007 (2018)
  [arXiv:1804.01120 [hep-th]].
  
  
\bibitem{Obied:2018sgi} 
  G.~Obied, H.~Ooguri, L.~Spodyneiko and C.~Vafa,
  ``De Sitter Space and the Swampland,''
  arXiv:1806.08362 [hep-th].
 		
\bibitem{Brandenberger:1988aj} 
  R.~H.~Brandenberger and C.~Vafa,
  ``Superstrings in the Early Universe,''
  Nucl.\ Phys.\ B {\bf 316}, 391 (1989).
		
 
\bibitem{Vafa:1996xn} 
  C.~Vafa,
  ``Evidence for F theory,''
  Nucl.\ Phys.\ B {\bf 469}, 403 (1996)
  [hep-th/9602022].
  
\bibitem{Beasley:2008dc} 
  C.~Beasley, J.~J.~Heckman and C.~Vafa,
  ``GUTs and Exceptional Branes in F-theory - I,''
  JHEP {\bf 0901}, 058 (2009)
  [arXiv:0802.3391 [hep-th]].
  
  	\end{thebibliography}
\end{document}